\documentclass[preprint,review, 12pt, authoryear]{elsarticle}

\usepackage{amssymb}

\usepackage[hyphens]{url}
\usepackage[breaklinks]{hyperref}   
\usepackage[hyphenbreaks]{breakurl}
\usepackage{longtable}
\usepackage{amsmath,amssymb,multirow,dcolumn,fancyhdr,charter,graphicx}
\usepackage{graphics}
\usepackage{xspace}
\usepackage{color,ulem,epstopdf}
\usepackage[extd]{journalnames} 
\usepackage{footnote}
\makesavenoteenv{tabular}
\makesavenoteenv{table}
\usepackage{setspace}
\usepackage{lineno}
\linenumbers

\begin{document}
\nolinenumbers
\begin{frontmatter}


\ead{malidib@nyu.edu}

\title{A machine-generated catalogue of Charon's craters and implications for the Kuiper belt}


 \author[as,aas]{Mohamad Ali-Dib}

\address[as]{Center for Astro, Particle and Planetary Physics (CAP$^3$), New York University Abu Dhabi, PO Box 129188, Abu Dhabi, UAE}
\address[aas]{Institut de recherche sur les exoplan\`etes, Universit\'e de Montr\'eal, 2900 boul. \'Edouard-Montpetit, Montr\'eal, H3T 1J4, Canada}

\end{frontmatter}

\newpage
\section*{Abstract}
In this paper we investigate Charon's craters size distribution using a deep learning model. This is motivated by the recent results of \cite{singer} who, using manual cataloging, found a change in the size distribution slope of craters smaller than 12 km in diameter, translating into a paucity of small Kuiper Belt objects. {These results were corroborated by \cite{robbins2021}, but {opposed} by \cite{morby}, necessitating an independent review.} Our MaskRCNN-based ensemble of models was trained on Lunar, Mercurian, and Martian crater catalogues and both optical and digital elevation images. We use a robust image augmentation scheme to force the model to generalize and transfer-learn into icy objects. With no prior bias  or exposure to Charon, our model  {find best fit slopes of $q=$-1.47$\pm$0.33  for craters smaller than 10 km, and $q=$-2.91$\pm$0.51 for craters larger than 15 km. These values indicate a clear change in slope around 15 km as suggested by \cite{singer} and thus independently confirm their conclusions. Our slopes however are both slightly flatter than those found more recently by  \cite{robbins2021}. Our trained models and relevant codes are available online on github.com/malidib/ACID . }

\section{Introduction}
Craters are the dominant morphological features on airless solar system bodies. Their frequency measurements are routinely used to infer the relative ages of different regions of an object, and absolute ages if calibration is possible \citep{murchie,pat}. They can moreover be compared between the apex and antapex regions of synchronous satellites to constrain whether the impacting debris had heliocentric or planetocentric orbits \citep{zahnle2}. 
Craters diameters distribution reflect the size distribution of the impactors, and that can be compared to known small bodies populations to constrain the nature of the asteroidal or cometary impactors reservoirs \citep{strom,dones,minton}. This is turn can be compared to different dynamical evolution models of the early solar system \citep{levison,zahnle}. On the other hand, the depth-to-Diameter distribution of craters holds valuable information on the crust's thickness and composition. \cite{schenk} for example used the d-to-D distribution slope on Europa to constrain its icy shell thickness. Finally, craters ellipticity distribution is directly related to the impact angles \citep{herrick}, and that can be used to constrain the obliquity evolution of the body \citep{holo}.

In the past few years, \textit{New Horizons} images allowed for the first detailed studies of the surfaces of Pluto and Charon.  
Using high resolution {LORRI (LOng Range Reconnaissance Imager, \cite{cheng2008}) and Ralph/MVIC imager \citep{reuter2008} data}, \cite{singer} investigated the size-frequency distributions of Pluto and Charon's craters. They concluded that this distribution has a {shallower slope} for craters smaller than $\sim$ 12 km, indicating a paucity in small craters and therefore {in small Trans-Neptunian Objects (TNOs), which is inconsistent with the main belt asteroids size distribution. {This was corroborated further by \cite{robbins2021} who generated an improved craters catalogue with manual counting, and came to the same conclusions.}
 {On the other hand, \cite{morby} combined the} \cite{singer} dataset with images of Arrokoth, a {cold classical Kuiper Belt Object} that was also imaged at medium resolution during \textit{New Horizons's} extended mission. Using the new dataset and a numerical model, \cite{morby} found a craters cumulative power law slope consistent with main belt asteroids.}

Identifying craters and refining  their size distributions has always been a manual process (as done for example by \cite{robbins} and \cite{singer}). This is cumbersome, specially for the small-radii end of the distribution where identification of craters becomes more time consuming. An automatized method has been proposed recently by \cite{ad5,ad4} who trained convolutional neural networks (CNNs) to {identify craters on the surface of the moon with a recall higher than 90\%.}  These algorithms were then successfully used to identify craters on Mercury \citep{ad5}, Mars \citep{lee}, as well as historic mining pits on Earth \citep{pits}. In this paper we investigate Charon's craters size-frequency distribution (SFD) using these automated techniques. {This allows us to have a robust and self-consistent catalogue of craters, with consistent size measurements.} In section \ref{methods} we present our machine learning model, and the results are discussed in section \ref{results}. We finally summarize and conclude in section \ref{summary}. 

\section{Methods}
\label{methods}
Our deep learning model of choice is the semantic segmentation framework MaskRCNN, \citep{he} \footnote{\url{https://github.com/facebookresearch/Detectron} and \url{https://github.com/matterport/Mask_RCNN}}. This algorithm has been successfully used by \cite{ad4} to constrain the size and ellipticity distribution of Lunar craters, and hence we refer the readers to that paper for more detailed background information on the framework. Here we discuss the main aspects of the model training and the differences with \cite{ad4}.  

\subsection{Training dataset}
The aim of this work is to investigate Charon's SFD using an unbiased model. For this end, we choose to train the model exclusively on a  dataset of Lunar, Mercurian, and Martian craters, containing digital elevation maps and optical images. The model is never exposed to icy objects at training or testing, {and is thus less prone to overfitting}. The massive size ($\sim$ 4 times larger than \cite{ad4}) of the dataset and its diversity that we increase furthermore using extensive image augmentation allows the final model to generalize very well through transfer learning to new objects, including Charon (this work), and comet 67P/Churyumov–Gerasimenko (future work). 

{Our images dataset is generated from the Moon's LRO/LOLA (Lunar Reconnaissance Orbiter/Lunar Orbiter Laser Altimeter) global DEM (Digital Elevation Map) and LRO/LROC WAC (wide Angle Camera) Global Optical Mosaic, Mercury's MESSENGER/MDIS (Mercury Dual Imaging System) Global Mosaic and Global DEM, and finally Mars' MGS/MOLA (Mars Global Surveyor/Mars Orbiter Laser Altimeter) DEM and Viking Global Mosaic V2 (we use Viking data instead of the newer THEMIS since it covers a larger area and it is simpler to use).  These are all available at \\ https://astrogeology.usgs.gov/search .\\ We supplemented the Lunar craters catalogues used by \cite{ad4} (\cite{Povilaitis} and \cite{Head}) with that of \cite{robbinsMoon} for smaller diameters. We used the \cite{robbinsMars} catalogue for Mars, and finally the \cite{mercurycat} catalogue for Mercury. The final dataset consists of 5411 ``poststamp'' 512$\times$512 pixel images that were generated through random cropping of the global maps, using the same procedure outlines in \cite{ad4}. {Random cropping allows us to generate on demand a practically infinite amount of unique training data from a finite map and catalogue.} {In our training set, as each individual catalogued crater is contained in a separate target mask, and thus considered as 1 training example, our dataset consists in total of 205618 training examples.} }

\subsection{Image augmentation} 

Our pipeline also includes extensive image augmentation to limit overfitting by increasing the diversity of the dataset in general, and  negate the directionally-biased effects of shadows in the optical images in particular. Our image augmentation routine is called during training when a poststamp image is loaded, and includes the following distinct effects:

\begin{itemize}
\item Image color inversion, { with 25\% probability of being applied to the image.} This allows the algorithm to detect craters no matter if they are darker or brighter than their immediate environment.  
\item Image rotation in multiples of 90 degrees, with a 33\% probability. This randomizes the direction of shadows, preventing any correlations between the incidence angle and detection efficiency.
\item Image flipping, with 33\% probability.
\item Adding Gaussian noise with a mean of 0 and standard-deviation of 0.7, with 17\% probability. This allows the model to generalize better to lower quality data. 
\item Adding ``salt and pepper'' noise with a pixel-alteration chance of 0.05, with 17\% probability. {The pixel-alteration chance, in addition to the mean and standard deviation of the Gaussian noise, were chosen through error and trial to add visible noise without overly distorting small craters. }
\item Random image perspective warping that adds a viewing angle to the image, instead of being orthogonal, with 17\% probability. This helps the algorithm detect highly elliptical craters. 
\end{itemize}
{Therefore, on average, any given image undergoes 1.5 random augmentations, with the median being 2. }

\subsection{Machine learning model}
Instead of relying on a single model, {we choose to implement an ensemble of models.} Ensembling is known to limit overfitting and improve model generalization \citep{ensemble,ensemble2}. Our ensemble is simply a collection of 14 MaskRCNN models with different parameters trained on the entirety or a subset of the data, each taken at a specific epoch {(how many
times to pass the input dataset through the network during training)}. The models parameters are summarized in Table 1.  We initially started by training a total of $\sim$ 100 models, and these 14 models were selected for having very low false positive rates when tested on a small (10 poststamps) random sample of Charon's images. The rationale is to have many models with low false positive rates even though for most of them the false negative rate is relatively high, as the ensemble of these models will have both a low false positive and negative rates. These rates were recalculated more precisely for Charon after the new catalogues have been generated, as discussed more in section \ref{inference}.

\begin{table}[]
\footnotesize

\begin{tabular}{llllll}
Model \# & epochs &loss weights & head only &  augmentation & data  \\
 \hline
1\footnote{Same model as \cite{ad4}}       & 13     & 1,1,1,1,1         & no         & no        & Moon only   \\
2       & 12      &  1,1,1,1,1 &  no         &   yes       & Moon only    \\
3      & 13      & 1,1,1,1,1  &   no        &   no  &   full dataset   \\
4      & 20      & 1,1,1,1,1&     no      &    rotation only      & Moon only      \\
5       & 18      &  1,1,1,1,1  &       no    &    no      &  full dataset  \\
6 \footnote{ \texttt{RPN\_ANCHOR\_SCALES} = (2, 4, 16, 64,128)  instead of (4, 8, 16, 32, 64), and \texttt{RNP\_NMS\_THRESOHOLD = 0.3
} instead of 0.7        }      & 7     & 1,1,0.1,0.1,1  &       no    &     no     &  full dataset    \\
7 \footnote{Training initially with loss weights = (1,1,1,1,1), then resumed with (1,4,1,1,1)}  & 40      &  (1,4,1,1,1)   &       no    &    no      &  full dataset    \\
8         & 255      & 1,1,1,1,1   &       no    &    no      &  full dataset   \\
9        & 450      &  1,1,1,1,1   &       no    &    no      &   full dataset   \\
10      & 153      &   1,1,1,1,1    &       no    &    no      & full dataset   \\
11\footnote{{head only: freeze (do not update the weights of) all layers except the dense (fully connected), non convolutional ``head'' layers.}}        & 135      & 1,1,1,1,1   &        yes   &    no      &  full dataset   \\
12        & 11      &  1,1,1,1,1  &       no    &    yes      & full dataset   \\
13       & 24      &  1,1,1,1,1 &       no    &   yes       &  full dataset   \\
14      & 18     &1,1,1,1,1   & no        & rotation only        &  full dataset   

\label{tablemodels}
\end{tabular}
\caption{Models ensemble parameters. Loss weights, are the weights given to the different (multi-)loss functions of MaskRCNN, and they are respectively \texttt{rpn\_class\_loss}, \texttt{rpn\_bbox\_loss}, \texttt{mrcnn\_class\_loss}, \texttt{mrcnn\_bbox\_loss}, \texttt{mrcnn\_mask\_loss}. These are all defined in \cite{he}, and explained in the Appendix below. All models were trained with the same learning rate of 10$^{-3}$. {``yes'' under augmentation means that the full augmentation routine was used. In models 1,2,4 only the Lunar data was used as it is the highest quality data available for training. }
{For models 8,9,10,11 we randomly subsampled the data, then train the models for 15 epochs for 8,9 and 9 epochs for 10,11. This process was repeated with new subsamples till reaching the quotes total number of epochs. } }
\end{table}

\subsection{Inference on Charon's images}
\label{inference}
\subsubsection{Methodology}
To generate Charon's craters catalogues, we use a sliding window on the New Horizons LORRI MVIC  300m/pixel \href{https://astrogeology.usgs.gov/search/map/Charon/NewHorizons/Charon_NewHorizons_Global_Mosaic_300m_Jul2017}{Mosaic} 
  to generate unique poststamp images on which {we make predictions using all 14 models}. {We however limit our predictions exclusively to the Vulcan Planitia region of Charon, and do not use the rest of the map (mainly Oz Terra). This is motivated by two reasons: 1- the planar nature of Vulcan Planitia facilitates craters identification and limits ambiguous surfaces, and 2- to maintain a reasonably homogeneous resolution across the map, as the imagery is not homogeneous with different regions and instruments having a variety of resolutions, incidence and emission angles, and terrain quality.}


  MaskRCNN models works best on image 512$\times$512 pixels, as this is the size of the training set images. However, we found that taking smaller poststamps (256$\times$256) then upscaling them to 512$\times$512 pixels improve the detection efficiency of small craters, with negligible effects on false positive rates. {The upscaling is done with \texttt{OpenCV}'s \texttt{resize} method, using bilinear interpolation .} Therefore, we generate catalogues using both 256$\times$256 and 512$\times$512 pixels sized sliding windows. In total, this result in 28 individual catalogues for the global map. We subsequently merge all of the catalogues, and filter out duplicates {by keeping track of the global coordinates for each detection}. 

{Duplicates filtration is done through calculating the intersection-over-union (IoU) of the bounding boxes of every pair combination, and removing those where the score is higher than {0.5 IoU (a value we found optimal by trial and error)}, also known as the Jaccard index, calculates the extent of overlap between the bounding boxes of objects pairs, then exclude one of them if the overlap is larger than a given parametric fraction. Note that two embedded craters where one is much larger than the other would have a small overlapping region (equal to the area of the smaller crater), and thus neither would be removed.} We finally end up with one definitive catalogue for the map that we use for to generate the SFD. For all inferences, we use a \texttt{DETECTION\_MIN\_CONFIDENCE} parameter\footnote{{MaskRCNN parameter defined as the confidence level threshold, above which the object will be classified. }} of 0.95, implying that it is a conservative catalogue.    

{At this stage we end up with a catalogue containing 1088 craters. We finally inspect the entire catalogue for false positives (non-crater identified as one) and false negatives (missed craters). We hence remove a total of 33 detections (or 3\% of the entire catalogue), and add 8 missed craters.  }


\section{Results}
\label{results}

{To study the craters size distribution, and for ease of comparison with \cite{singer}, we choose to follow them in representing the SFD as an R-plot. In these plots the The differential number of craters per diameter ($dN/dD \propto D^q$ where q is the slope) is divided by $D^{-3}$. 
To calculate the slopes below we fit the R-plots directly, then subtract 3 to obtain the differential distribution slope $q$. }

{Our  automatically generated but {hand-curated} craters catalogue is shown in Fig. \ref{fig:charonGlobal} and Charon's R-plot is shown in \ref{fig:charon1}. Visually inspecting Fig. \ref{fig:charon1}, a change in the slope between 10 and 15 km is visible. {Our {slopes} are also moderately consistent with the catalogue of \cite{robbins2021}, also plotted. Using $\chi^2$ minimization, we find best fit slopes for $q$ of -1.47$\pm$ 0.33  for craters smaller than 10 km, and -2.91$\pm$ 0.51 for craters larger than 15 km. {Both of our slopes are slightly shallower than those found by \cite{robbins2021} who found $q\sim -3.9 \pm 0.6$ for large craters and $q\sim -2.0 \pm 0.2$ for small ones in the Vulcan Planitia region. Our slopes are however consistent with those found for some Pluto regions, where \cite{robbins2021} reported for the mid-latitude strip $q\sim -2.4 \pm 0.5$ and $q\sim -1.9 \pm 0.5$ for large and small craters respectively. }}

We verify this change in slope using Bayesian information criterion (BIC) analysis. {BIC analysis is used in models selection by assigning a coefficient value to each model. The actual values have no real meaning. However, models with smaller coefficients (where negative numbers are smaller than positive ones), are preferred over models with larger coefficients}. Our analysis needs to account for two properties of the data: the error bars, and the fact that small craters are much more common than large ones (this is true in both cumulative and non-cumulative SFDs). Our procedure hence consists of splitting the data at 10-15 km diameter, then calculating the BICs for two models with $q$ = -2.91 and $q$ = -1.47 while accounting for the uncertainties. This is done by replacing each individual data point with one sampled randomly from its uncertainty interval, then calculating the average BIC of 10$^4$ repetitions. Splitting the data prevents small craters from dominating the BIC no matter what slope is preferred for the far less numerous larger ones. We get final BIC values for craters larger than 15 km of 89.28 for $q$ = -2.91, and 177.20 for $q$ = -1.47. Therefore, for large craters, the $q$ = -2.91 model is strongly preferred over $q$ = -1.47. Considering small craters we get a BIC of 223.35 for  $q$ = -2.91 and 161.66 for $q$ = -1.47, implying that $q$ = -1.47 is strongly preferred over $q$ = -2.91 for craters smaller than 10 km. Bayesian analysis hence favors a change in Charon's SFD slope between 10 and 15 km diameter.  Our results strengthen the conclusions of \cite{singer}, as this slope change translates into a paucity of impactors smaller than 1.1 km.

\begin{figure*}
	\begin{center}
		{\includegraphics[scale=0.30]{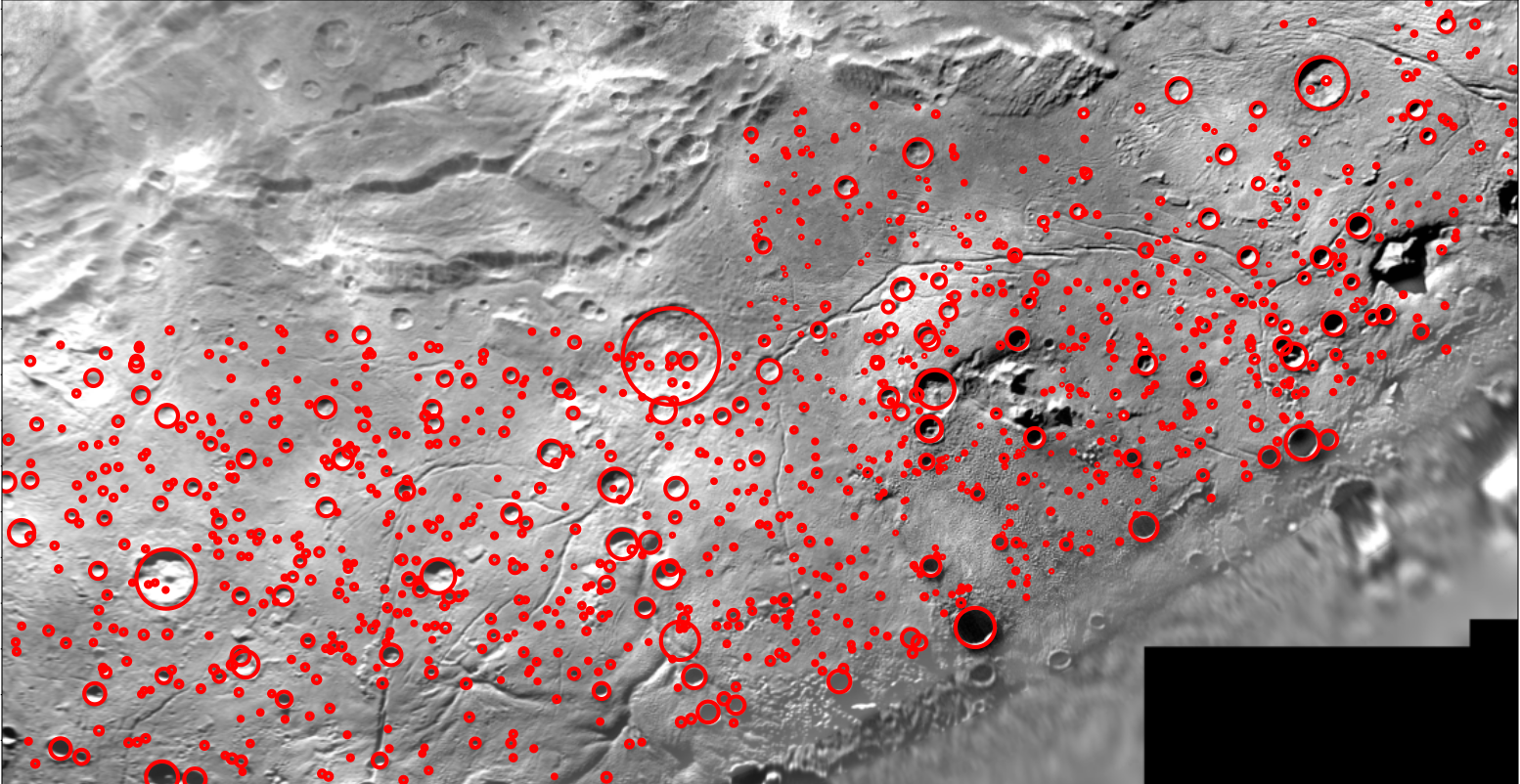}}
		\caption{{A map of Charon's Vulcan Planitia region, with the final craters catalogue plotted.  }  }
		\label{fig:charonGlobal}
	\end{center}
\end{figure*}

\begin{figure*}
	\begin{center}
		{\includegraphics[scale=0.42]{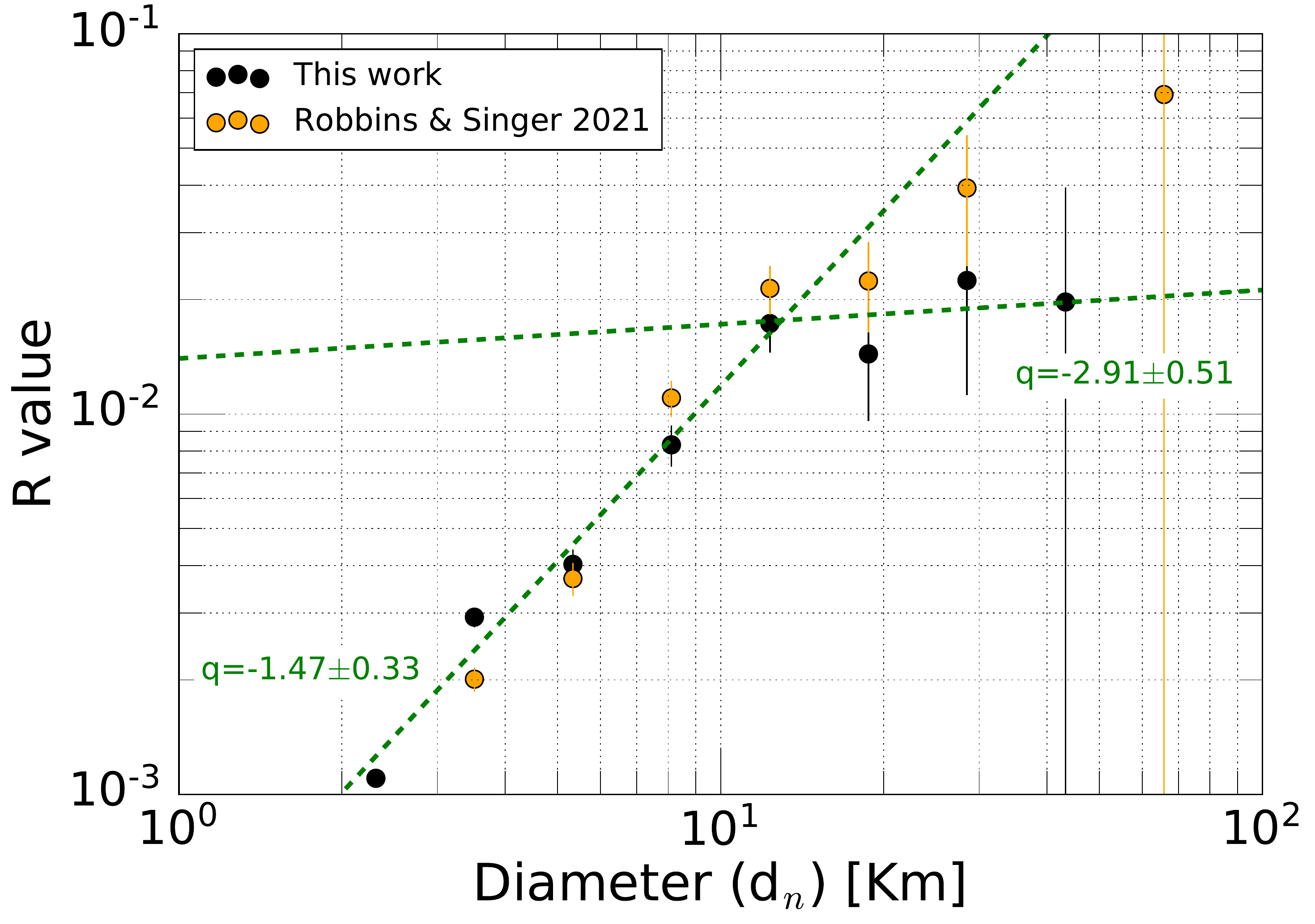}}
		\caption{Charon's R-plot (differential size-frequency distribution normalized to D$^{-3}$), compared to \cite{robbins2021}. Both curves show a change in slope {between 10 and 15 km}. {Note that in R-plots a horizontal line corresponds to q=-3.} }
		\label{fig:charon1}
	\end{center}
\end{figure*}


\section{Summary \& Conclusions}
\label{summary}
In this paper we investigated the crater size-frequency distribution of the Vulcan Planitia region of Charon using a MaskRCNN deep learning ensemble of 14 models. After training the model on a plethora of rocky objects images and catalogues, and augmenting the dataset through random rotation, flipping, perspective change, and noise addition, we obtain a model capable of transfer-learning to icy objects. Applying the model to Charon, we independently confirm the results of \cite{singer}  and \cite{robbins2021} who found a change of slope {between 10 and 15 km} diameter, implying a paucity of KBOs smaller than $\sim$ 1.1 km. {The slopes we found are however slightly shallower than \cite{robbins2021}.} We do not investigate Pluto's craters, as we found that its more complex and rugged terrain poses challenges for our model. 

\subsection*{Implications for future solar system missions}
{Multiple} other major icy objects in the solar system have been partially or near completely mapped at kilometre or sub-km resolutions. The amount of information present in these images, in addition to complementary measurements, is overwhelming. 
This will only get exacerbated with upcoming missions providing higher resolution imagery. The Europa Clipper's instrument EIS for example will map Europa's surface at a maximum resolution of 50 m, down to possibly 0.5 m \citep{turtle}. JUICE's JANUS camera on the other hand will map Ganymede and parts of Callisto with a resolution of 400 m, down to possibly 2.4 m \citep{pal}. Properly analyzing the existent amount of data is already challenging, so one can quickly see that new techniques altogether will be necessary for the Clipper and JUICE. {Fortunately, computer vision techniques such as the one used in this work proved to be incredibly powerful tools for planetary surfaces objects identification.} These techniques should be developed and tested on Jupiter and Saturn moons with published catalogues, in preparation for the upcoming data stream.

\section*{Acknowledgments}
We thank S. Robbins and an anonymous referee for their useful comments that helped
us improve our manuscript significantly. This work was supported by Tamkeen under the NYU Abu Dhabi Research Institute grant CAP3, and M.A.-D. was supported through a CAP3 fellowship.

\section*{Appendix}
\subsection*{Some MaskRCNN terminology}
\texttt{rpn\_class\_loss}: loss function accounting for the improper classification of anchor boxes. It is higher when multiple objects are not detected.

\texttt{rpn\_bbox\_loss}: loss function accounting for the localization accuracy. It is higher if objects are detected but the bounding boxes are incorrect. 

\texttt{mrcnn\_class\_loss}: loss function that increases if the object is detected but classified incorrectly.  

\texttt{mrcnn\_bbox\_loss}: loss function that increases if the object is correctly classified, but its localization is imprecise. 

 \texttt{mrcnn\_mask\_loss}: loss function that increases if the masks at the pixel level are not accurate. 



\end{document}